\documentstyle[12pt,epsf]{article}

\catcode`\@=11
\long\def\@makefntext#1{ 
\protect\noindent \hbox to 3.2pt {\hskip-.9pt  
$^{{\ninerm\@thefnmark}}$\hfil}#1\hfill} 

\def\thefootnote{\fnsymbol{footnote}}
 \def\@makefnmark{\hbox to 0pt{$^{\@thefnmark}$\hss}}  
        
\def\ps@myheadings{\let\@mkboth\@gobbletwo
\def\@oddhead{\hbox{} 
\rightmark\hfil\ninerm\thepage}   
\def\@oddfoot{}\def\@evenhead{\ninerm\thepage\hfil 
\leftmark\hbox{}}\def\@evenfoot{}
\def\sectionmark##1{}\def\subsectionmark##1{}}

\begin{document}

\newcommand{\symbolfootnote}{\renewcommand{\thefootnote}
        {\fnsymbol{footnote}}}
\renewcommand{\thefootnote}{\fnsymbol{footnote}}
\newcommand{\alphfootnote}
        {\setcounter{footnote}{0}
         \renewcommand{\thefootnote}{\sevenrm\alph{footnote}}}

\newcounter{sectionc}\newcounter{subsectionc}\newcounter{subsubsectionc}
\renewcommand{\section}[1] {\vspace{0.6cm}\addtocounter{sectionc}{1} 
\setcounter{subsectionc}{0}\setcounter{subsubsectionc}{0}\noindent 
        {\bf\thesectionc. #1}\par\vspace{0.4cm}}
\renewcommand{\subsection}[1] {\vspace{0.6cm}\addtocounter{subsectionc}{1} 
        \setcounter{subsubsectionc}{0}\noindent 
        {\it\thesectionc.\thesubsectionc. #1}\par\vspace{0.4cm}}
\renewcommand{\subsubsection}[1] {\vspace{0.6cm}\addtocounter{subsubsectionc}{1}
        \noindent {\rm\thesectionc.\thesubsectionc.\thesubsubsectionc. 
        #1}\par\vspace{0.4cm}}
\newcommand{\nonumsection}[1] {\vspace{0.6cm}\noindent{\bf #1}
        \par\vspace{0.4cm}}
                                                 
\newcounter{appendixc}
\newcounter{subappendixc}[appendixc]
\newcounter{subsubappendixc}[subappendixc]
\renewcommand{\thesubappendixc}{\Alph{appendixc}.\arabic{subappendixc}}
\renewcommand{\thesubsubappendixc}
        {\Alph{appendixc}.\arabic{subappendixc}.\arabic{subsubappendixc}}

\renewcommand{\appendix}[1] {\vspace{0.6cm}
        \refstepcounter{appendixc}
        \setcounter{figure}{0}
        \setcounter{table}{0}
        \setcounter{equation}{0}
        \renewcommand{\thefigure}{\Alph{appendixc}.\arabic{figure}}
        \renewcommand{\thetable}{\Alph{appendixc}.\arabic{table}}
        \renewcommand{\theappendixc}{\Alph{appendixc}}
        \renewcommand{\theequation}{\Alph{appendixc}.\arabic{equation}}
        \noindent{\bf Appendix \theappendixc #1}\par\vspace{0.4cm}}
\newcommand{\subappendix}[1] {\vspace{0.6cm}
        \refstepcounter{subappendixc}
        \noindent{\bf Appendix \thesubappendixc. #1}\par\vspace{0.4cm}}
\newcommand{\subsubappendix}[1] {\vspace{0.6cm}
        \refstepcounter{subsubappendixc}
        \noindent{\it Appendix \thesubsubappendixc. #1}
        \par\vspace{0.4cm}}

\def\abstracts#1{{
        \centering{\begin{minipage}{30pc}\tenrm\baselineskip=12pt\noindent
        \centerline{\tenrm ABSTRACT}\vspace{0.3cm}
        \parindent=0pt #1
        \end{minipage} }\par}} 

\newcommand{\bibit}{\it}
\newcommand{\bibbf}{\bf}
\renewenvironment{thebibliography}[1]
        {\begin{list}{\arabic{enumi}.}
        {\usecounter{enumi}\setlength{\parsep}{0pt}
\setlength{\leftmargin 1.25cm}{\rightmargin 0pt}
         \setlength{\itemsep}{0pt} \settowidth
        {\labelwidth}{#1.}\sloppy}}{\end{list}}

\topsep=0in\parsep=0in\itemsep=0in
\parindent=1.5pc

\newcounter{itemlistc}
\newcounter{romanlistc}
\newcounter{alphlistc}
\newcounter{arabiclistc}
\newenvironment{itemlist}
        {\setcounter{itemlistc}{0}
         \begin{list}{$\bullet$}
        {\usecounter{itemlistc}
         \setlength{\parsep}{0pt}
         \setlength{\itemsep}{0pt}}}{\end{list}}

\newenvironment{romanlist}
        {\setcounter{romanlistc}{0}
         \begin{list}{$($\roman{romanlistc}$)$}
        {\usecounter{romanlistc}
         \setlength{\parsep}{0pt}
         \setlength{\itemsep}{0pt}}}{\end{list}}

\newenvironment{alphlist}
        {\setcounter{alphlistc}{0}
         \begin{list}{$($\alph{alphlistc}$)$}
        {\usecounter{alphlistc}
         \setlength{\parsep}{0pt}
         \setlength{\itemsep}{0pt}}}{\end{list}}

\newenvironment{arabiclist}
        {\setcounter{arabiclistc}{0}
         \begin{list}{\arabic{arabiclistc}}
        {\usecounter{arabiclistc}
         \setlength{\parsep}{0pt}
         \setlength{\itemsep}{0pt}}}{\end{list}}

\newcommand{\fcaption}[1]{
        \refstepcounter{figure}
        \setbox\@tempboxa = \hbox{\tenrm Fig.~\thefigure. #1}
        \ifdim \wd\@tempboxa > 6in
           {\begin{center}
        \parbox{6in}{\tenrm\baselineskip=12pt Fig.~\thefigure. #1 }
            \end{center}}
        \else
             {\begin{center}
             {\tenrm Fig.~\thefigure. #1}
              \end{center}}
        \fi}

\newcommand{\tcaption}[1]{
        \refstepcounter{table}
        \setbox\@tempboxa = \hbox{\tenrm Table~\thetable. #1}
        \ifdim \wd\@tempboxa > 6in
           {\begin{center}
        \parbox{6in}{\tenrm\baselineskip=12pt Table~\thetable. #1 }
            \end{center}}
        \else
             {\begin{center}
             {\tenrm Table~\thetable. #1}
              \end{center}}
        \fi}

\def\@citex[#1]#2{\if@filesw\immediate\write\@auxout
        {\string\citation{#2}}\fi
\def\@citea{}\@cite{\@for\@citeb:=#2\do
        {\@citea\def\@citea{,}\@ifundefined
        {b@\@citeb}{{\bf ?}\@warning
        {Citation `\@citeb' on page \thepage \space undefined}}
        {\csname b@\@citeb\endcsname}}}{#1}}

\newif\if@cghi
\def\cite{\@cghitrue\@ifnextchar [{\@tempswatrue
        \@citex}{\@tempswafalse\@citex[]}}
\def\citelow{\@cghifalse\@ifnextchar [{\@tempswatrue
        \@citex}{\@tempswafalse\@citex[]}}
\def\@cite#1#2{{$\null^{#1}$\if@tempswa\typeout
        {IJCGA warning: optional citation argument 
        ignored: `#2'} \fi}}
\newcommand{\citeup}{\cite}

\def\fnm#1{$^{\mbox{\scriptsize #1}}$}
\def\fnt#1#2{\footnotetext{\kern-.3em
        {$^{\mbox{\sevenrm #1}}$}{#2}}}

\font\twelvebf=cmbx10 scaled\magstep 1
\font\twelverm=cmr10 scaled\magstep 1
\font\twelveit=cmti10 scaled\magstep 1
\font\elevenbfit=cmbxti10 scaled\magstephalf
\font\elevenbf=cmbx10 scaled\magstephalf
\font\elevenrm=cmr10 scaled\magstephalf
\font\elevenit=cmti10 scaled\magstephalf
\font\bfit=cmbxti10
\font\tenbf=cmbx10
\font\tenrm=cmr10
\font\tenit=cmti10
\font\ninebf=cmbx9
\font\ninerm=cmr9
\font\nineit=cmti9
\font\eightbf=cmbx8
\font\eightrm=cmr8
\font\eightit=cmti8

\begin{flushright}
TIT/HEP--356 \\
November, 1996 
\end{flushright}
\vspace{2mm}
\centerline{\tenbf SUPERSYMMETRIC YANG-MILLS THEORIES 
}
\baselineskip=16pt
\centerline{\tenbf IN $1 + 1$ DIMENSIONS
}
\vspace{0.8cm}
\centerline{\tenrm 
 Norisuke Sakai \footnote{
\tt e-mail: nsakai@th.phys.titech.ac.jp} 
}
\baselineskip=13pt
\centerline{\tenit Department of Physics, Tokyo Institute of Technology, }
\baselineskip=12pt
\centerline{\tenit Oh-okayama, Meguro, Tokyo 152, Japan}
\vspace{0.9cm}
\abstracts{
Supersymmetric Yang-Mills theories are considered in $1+1$ dimensions. 
Firstly physical mass spectra of supersymmetric Yang-Mills theories in 
$1+1$ dimensions are evaluated in the light-cone gauge with a 
compact spatial dimension. 
The supercharges are constructed in order to provide a manifestly 
supersymmetric infrared regularization 
for the discretized light-cone approach. 
By exactly 
diagonalizing the supercharge matrix between up to 
several hundred color singlet bound states, we find a rapidly increasing 
density of states as mass increases. 
Interpreting this limiting density of states as the stringbehavior, 
we obtain the Hagedron temperature 
$\beta_H=0.676 \sqrt{\pi \over g^2 N}$. 
Secondly we have examined the vacuum structure of supersymmetric 
Yang-Mills theories in $1+1$ dimensions. 
SUSY allows only periodic boundary conditions for both fermions 
and bosons. 
By using the Born-Oppenheimer approximation for the weak coupling 
limit, we find that the vacuum energy vanishes, and hence the SUSY 
is unbroken. 
Other boundary conditions are also studied. 
The first part is based on a work in collaboration with Y. Matsumura and 
T. Sakai. 
The second part is based on a work in collaboration with H. Oda and T. Sakai. 
}

\vfil
\twelverm   
\baselineskip=14pt
\section{Introduction
}
\vglue 1pt
Supersymmetric theories have now become standard models for the 
unified theory. 
Both as a model for grand unified theories and as a low energy 
effective theory for superstring, the dynamics of supersymmetric 
Yang-Mills 
theories is a fascinating subject. 
The most outstanding problem in unified theories based on 
supersymmetry is to understand the supersymmetry breaking. 
Nonperturbative dynamics is expected to be essential to study the 
mechanisms of supersymmetry breaking. 

One of the most popular models for the supersymmetry breaking 
is currently to assume the gaugino bilinear condensation in 
the supersymmetric Yang-Mills theories  \cite{Nilles}. 
Although the condensation itself may not break supersymmetry 
in the supersymmetric gauge theories, it will give rise to 
the supersymmetry breaking if embedded in supergravity \cite{VeYa}. 
Since the fermion bilinear condensation is implied by 
 the chiral symmetry breaking in QCD, one can expect a similar
nonperturbative effects in supersymmetric Yang-Mills theories. 
Moreover, recent progress in understanding duality in supersymmetric 
Yang-Mills theories opened up a rich arena for studying the 
nonperturbative effects in supersymmetric gauge 
theories \cite{SeWi}. 

It has been quite fruitful to study Yang-Mills theories in $1+1$ 
dimensions 
instead of studying directly the four dimensional counterpart. 
In $1+1$ dimensions, gauge field itself has no dynamical 
degree of 
freedom as a field theory, but gives rise to a confining potential 
for colored particles \cite{THooft}. 
Many aspects of color singlet bound states can be explored by solving 
the theory in the large $N$ limit \cite{CaCoGr}. 
Unfortunately the supersymmetric gauge multiplet contains genuine 
dynamical degree of freedom in the adjoint representation of the 
gauge group contrary to ordinary Yang-Mills theory \cite{Ferrara}. 
Therefore one cannot obtain a simple closed form for the color singlet 
bound states even in the large $N$ limit. 

There has been progress in studying the dynamics of matter fields in the 
adjoint representation in ordinary Yang-Mills  
theories \cite{DaKl}. 
They have used the light-cone quantization and compactified the 
spatial dimension to give discrete momenta. 
In this discretized light-cone quantization approach, 
one can diagonalize the mass matrix for finite number of light-cone 
momenta and can hope to obtain the infinite volume limit 
eventually \cite{PuBr} \cite{HoBr}. 

More recently, gauge theories in $1+1$ dimensions 
with matter in adjoint representations 
was studied focusing attention on zero modes \cite{LeShTh}. 
The zero modes are generally important in revealing nontrivial 
vacuum structures such as the vacuum condensate 
\cite{MaYa} \cite{KoSaSa}.

The Born-Oppenheimer approximation in the weak coupling region 
has been used to study the vacuum structure of gauge theories 
with adjoint fermions \cite{Lenz}. 
Since the gauge coupling in $1+1$ dimensions has the dimension of 
mass, the weak coupling is characterized by 
\begin{equation}
gL \ll 1 ,
\end{equation}
where $L$ is the interval of the compactified spatial dimension. 
The fermion bilinear was found to possess a nonvanishing vacuum 
expectation value which exhibits instanton-like dependence 
on gauge coupling. 
The Yang-Mills theories with adjoint 
fermions were also studied at finite temperature and 
were shown to be dominated by instanton effects at high temperatures 
\cite{Smilga}. 
The Born-Oppenheimer approximation has been used to study SUSY
gauge theories in four dimensions \cite{Wit} \cite{Sm}.

In spite of these investigations of Yang-Mills gauge theories with 
adjoint scalar and spinor matter fields, there are two points 
which necessitate a new analysis of physical spectra 
in the case of supersymmetric 
gauge theories. 
The first point is that the coexistence of spinor and scalar 
gives rise to a large number of new ``mixed'' physical states, 
partly consisting of spinors and partly of scalars as 
constituents. 
The second point is the presence of a specific amount of 
the Yukawa interaction 
which is a distinguishing feature of the supersymmetric Yang-Mills 
theory \cite{Ferrara}. 

As for the possibility of the SUSY breaking, the Witten index \cite{Wit}  
of the SUSY Yang-Mills theories 
has been calculated recently, and was found to be nonvanishing 
in $1 + 1$ dimensions \cite{Li}. 
Although this result implies no possibility for spontaneous SUSY 
breaking, we feel it still worthwhile to study the vacuum of the SUSY 
Yang-Mills theories in $1+1$ dimensions by a more detailed dynamical 
calculation, since the calculation of the Witten index involved a certain 
regularization of bosonic zero modes which may not be easily 
justified. 

In view of this situation, we have studied SUSY Yang-Mills theories 
from two perspectives: 
mass specra and vacuum structures. 
The mass spectra has been computed in the discretized light-cone 
quantization in collaboration with Y. Matsumura and T. Sakai, 
\cite{MSS}
and the vacuum structure is studied by the Born-Oppenheimer 
approxmation in collaboration with H. Oda and T. Sakai 
\cite{OSS}.

We construct the supercharge explicitly and specify an 
infrared regularization for supercharge by means of 
the discretized version of the principal value prescription. 
By using the supercharge, we succeed in overcoming ambiguities 
in prescribing the infrared regularization for the light-cone 
Hamiltonian. 
As a result, the regularization preserves the supersymmetry algebra 
manifestly. 
For light-cone momenta 
up to 8 units of the smallest momentum, 
we find several hundred color singlet bound states of bosons and the 
same number of fermions. 
We exactly diagonalize the supercharge instead of the Hamiltonian to 
obtain masses, degeneracies, and the average number of constituents 
in these bound states. 
We observe that the density of the bound states as a function of their 
masses tends to converge in the large volume limit. 
It is consistent with the rapidly increasing density of states 
suggested by the closed string interpretation. 
Since we preserve supersymmetry at each stage of 
our study, we naturally obtain exact correspondence 
between bosonic and fermionic color singlet bound states. 
We have also introduced a mass term for adjoint scalar and/or spinor fiels. 
Since these fields are superpartner of gauge field which are strictly 
massless, these terms break supersymmetry softly. 
we find indeed that the degeneracy of the mass spectra for color singlet 
bosons and fermions is lifted. 

Light-cone approach is notoriously difficult to unravel the vacuum structure. 
We need alternative systematic 
approaches to study the vacuum structure. 
To study the vacuum structure, 
we use the Born-Oppenheimer approximation in the weak coupling 
region. 
To formulate the weak coupling limit, we need to 
compactify the spatial direction. 
Since gauge fields naturally follow periodic boundary conditions, 
we need to require the same periodic boundary conditions for scalar 
and spinor fields in order not to break SUSY by hand. 
We have found that the ground state has a vanishing vacuum energy, 
suggesting that SUSY is not broken spontaneously. 
This result is consistent with the result on the Witten index \cite{Li}. 
We also examine all four possibilities of periodic and anti-periodic 
boundary conditions for fermions and bosons \cite{OSS}. 

\twelverm   
\baselineskip=14pt
\section{
SUSY Yang-Mills Theories in $1+1$ Dimensions
}
\vglue 1pt
In two-dimensions, the gauge field $A^\mu$ is contained in a 
supersymmetric multiplet consisting of a Majorana fermion $\Psi$ 
and a scalar $\phi$ in the adjoint representation of the gauge 
group together with gauge field itself \cite{Ferrara}. 
After choosing the Wess-Zumino gauge, we have an 
action 
\begin{equation}
S =
\int d^2 x \, {\rm tr} \Bigg[-{1\over 4 g^{2}} F_{\mu \nu}F^{\mu \nu}
+\frac{1}{2} D_{\mu}\phi D^{\mu}\phi 
+i\bar{\Psi} \gamma^{\mu} D_{\mu} \Psi \nonumber 
  -2ig\phi\bar{\Psi} \gamma_{5} \Psi \Bigg], 
\label{eq:sacda}
\end{equation}
where 
$A_{\mu}$, 
$\phi$, 
$\Psi$, 
and 
$\bar \Psi=\Psi^T \gamma^0$ 
are traceless $N\times N$ hermitian matrix for $U(N)$ ($SU(N)$) gauge 
group, 
 $g$ is the gauge coupling constant,  
$F_{\mu\nu} = \partial_{\mu}A_{\nu} - \partial_{\nu}A_{\mu}+i[A_{\mu} 
A_{\nu}]$ and $D_{\mu}$ is the usual covariant derivative 
\begin{equation}
D_{\mu}\phi = \partial_{\mu}\phi+i[A_{\mu}, \phi], \quad 
D_{\mu}\Psi = \partial_{\mu}\Psi+i[A_{\mu}, \Psi] . 
\end{equation}
The supersymmetry dictates the presence of the Yukawa type interaction 
between the adjoint spinor and scalar fields with the strength of 
the gauge coupling. 
The supersymmetric Yang-Mills gauge theory in two-dimensions 
can be obtained by a dimensional reduction from the supersymmetric 
Yang-Mills gauge theory in three dimensions. 
The adjoint scalar field can be understood as the component of the gauge 
field in the compactified dimension and the Yukawa coupling is 
nothing but the gauge interaction in this compactified extra dimension. 

In the Wess-Zumino gauge, the remaining invariances 
of the action are the usual gauge invariance and a supertransformation 
which is obtained by combining the supertransformation 
and the compensating gauge transformation in the superfield 
formalism. 
The corresponding spinor supercurrent $j^\mu$ is given by 
\begin{equation}
\bar \epsilon j^\mu = 
\, {\rm tr}\left[-\sqrt{2}\bar \epsilon \Psi D^\mu \phi+i{1 \over \sqrt{2}g}
 \epsilon^{\nu\lambda}F_{\nu\lambda} 
 \bar \epsilon \gamma^\mu \Psi+\sqrt{2} \bar \epsilon \gamma_5 
\Psi \epsilon^{\mu\nu} D_\nu \phi\right] .
\label{supercurrent}
\end{equation}

We introduce the light-cone coordinates where the line 
element $ds^2$ is given by 
\begin{equation}
x^{\pm}=\frac{1}{\sqrt{2}}(x^{0}\pm x^{1}), 
\quad 
ds^2=(dx^0)^2-(dx^1)^2=2dx^+dx^-. 
\end{equation}
We decompose the spinor and use gamma matrices 
\begin{equation}
\Psi_{ij}=2^{-1/4} (\psi_{ij},\chi_{ij})^{T}, 
\qquad 
\gamma^0 =\sigma_2,\; \gamma^1 =i\sigma_1,\;
\gamma_5 = \gamma^0 \gamma^1=\sigma_3.
\label{gammamatrix}
\end{equation}
Taking the light-cone gauge 
$
A_{-} = A^{+} =0 
$ and $x^+$ as time, we find the action 
\begin{eqnarray}
S &\!\!\!=&\!\!\! 
\int dx^{+}dx^{-} \, {\rm tr} \Bigg[\partial_{+}\phi\partial_{-}\phi
+i\psi\partial_{+}\psi +i\chi\partial_{-}\chi
\nonumber \\
&\!\!\!&\!\!\!
+\frac{1}{2g^2} (\partial_{-}A_{+})^2 
+A_{+}J^{+} 
+\sqrt{2}g\phi\{\psi,\chi\}
 \Bigg], 
\label{eq:action}
\end{eqnarray} 
where the current $J^+$ 
receives 
contributions from the scalar $J_{\phi}^{+}$ and the spinor 
$J_{\psi}^{+}$ 
\begin{equation}
J^+=J_{\phi}^{+} + 
J_{\psi}^{+}, \quad 
J_{\phi}^{+} = i[\phi,\partial_{-}\phi], \quad 
J_{\psi}^{+} = 2\psi\psi.
\label{totalcurrent}
\end{equation}

We do not need Faddeev-Popov ghosts 
in this gauge. 
Since the action contains no time derivative for the 
gauge potential $A_{+}$ and the left-moving fermion $\chi$, 
they can be eliminated by means of constraints obtained as 
their Euler-Lagrange equations 
\begin{equation}
i\sqrt{2} \partial_{-}\chi 
- g[\phi,\psi] = 0,
\quad
\partial_{-}^{2} \bar{A}_{+} - g^2 J^{+} = 0.
\label{eq:const}
\end{equation}
where $\bar{A}_{+}$ is the non-zero mode of $A_{+}$.
The zero mode of $A_{+}$ plays 
the role of a Lagrange multiplier which provides a constraint
\begin{equation} 
\int dx^- J^+ =0.
\label{shiki2}
\end{equation}   
This constraint will give a restriction for physical states in quantum 
theory.
After eliminating the fields $A_+$ and $\chi$, 
we find that the action becomes 
\begin{equation}
S = \int dx^{+}dx^{-} \, {\rm tr} \Bigg[  \partial_{+}\phi\partial_{-}\phi
+i\psi\partial_{+}\psi 
 +\frac{g^2}{2}J^{+}\frac{1}{\partial_{-}^{2}}J^{+} 
-\frac{1}{2}ig^2 [\phi,\psi]\frac{1}{\partial_{-}} [\phi,\psi] \Bigg]. 
\label{eq:actionphys}
\end{equation} 
Let us note that the constraints give rise to non-local terms 
in the action. 

By the Noether procedure, we construct the energy momentum 
tensor $T^{\mu\nu}$, and light-cone momentum and energy 
$P^{\pm}= \int dx^{-} T^{+\pm}$ on a constant light-cone time 
\begin{equation}
P^{+} =  \int dx^{-}
\, {\rm tr}\biggl[(\partial_{-}\phi)^{2} +i\psi\partial_{-}\psi \biggl], 
\label{eq:p+}
\end{equation}
\begin{equation}
P^{-} =  \int dx^{-}
\, {\rm tr} \Bigg[- \frac{g^2}{2}  J^{+} \frac{1}{\partial_{-}^{2}} J^{+}
+\frac{i}{2} g^2 [\phi,\psi]\frac{1}{\partial_{-}}[\phi,\psi] \Bigg].
\label{eq:p-}
\end{equation}

The supercharges $Q_1$ and $Q_2$ 
are defined as integrals of the upper and 
lower components of the spinor supercurrent $j^\mu=(j_1^\mu, j_2^\mu)$ 
in eq.(\ref{supercurrent}) 
\begin{equation}
Q_1 \equiv \int dx^- j_1^+ 
=
 2^{1/4}\int dx^{-}\, {\rm tr}\left[\phi\partial_{-}\psi-\psi\partial_{-}\phi
\right] ,
\end{equation}
\begin{eqnarray}
Q_2 
&\!\!\!
\equiv 
&\!\!\!
\int dx^- j_2^+ 
=
 2^{3/4}g\int dx^{-}\, {\rm tr}\left[J^+{1 \over \partial_-}\psi\right] 
 \nonumber \\[.5mm]
&\!\!\!
=
&\!\!\!
 2^{3/4}g\int dx^{-}\, {\rm tr}
\left\{\left(i[\phi, \partial_-\phi] + 2\psi\psi\right)
{1 \over \partial_-}\psi\right\} . 
\end{eqnarray}

Using the conjugate momenta 
$\pi_{\phi}= \partial {\cal L}/\partial (\partial_{+}\phi)
= \partial_{-}\phi$ 
for adjoint scalar field $\phi_{ij}$ 
and $\pi_{\psi}= \partial {\cal L}/\partial (\partial_{+}\psi)
= i\psi$
for adjoint spinor field $\psi_{ij}$, 
the canonical (anti)commutation relation 
are given at equal light-cone time $x^{+} = y^{+}$ by 
\begin{equation}
[\phi_{ij}(x),\partial_{-}\phi_{kl}(y)] 
=
i \{\psi_{ij}(x), \psi_{kl}(y)\}   
=
\frac{1}{2}i \delta(x^{-}-y^{-})\delta_{il} \delta_{jk}.
\label{cancomrel}
\end{equation}

We expand the fields 
in modes with momentum $k^+$ 
at light-cone time $x^+=0$ 
\begin{equation}
\phi_{ij}(x^{-} ,0) =  \frac{1}{\sqrt{2\pi}} \int_{0}^{\infty}
\frac{dk^{+}}{\sqrt{2k^{+}}} 
\left( a_{ij}(k^{+}){\rm e}^{-ik^{+}x^{-}} +
a_{ji}^{\dag}(k^{+}) {\rm e}^{ik^{+}x^{-}}\right) ,
\label{eq:modephi}
\end{equation}
\begin{equation}
\psi_{ij}(x^{-} ,0) =  \frac{1}{2\sqrt{\pi}} \int_{0}^{\infty} dk^{+}
\left (b_{ij}(k^{+}){\rm e}^{-ik^{+}x^{-}} +
b_{ji}^{\dag}(k^{+}){\rm e}^{ik^{+}x^{-}}\right ). 
\label{eq:modepsi}
\end{equation}
The canonical (anti-)commutation relations (\ref{cancomrel}) 
are satisfied by 
\begin{equation}
[a_{ij}(k^{+}),a_{lk}^{\dag}(\tilde{k}^{+})]   = 
\{b_{ij}(k^{+}), b_{lk}^{\dag}(\tilde{k}^{+})\}   = 
\delta(k^{+} - \tilde{k}^{+}) \delta_{il} \delta_{jk}. 
\end{equation}

\twelverm   
\baselineskip=14pt
\section{
Discretized Light-Cone Quantization of Superchage
}
\vglue 1pt
In order to prescribe the infrared regularization precisely and 
to evaluate the mass spectrum in spaces with finite number of physical 
states, we compactify spatial direction $x^{-}$ 
to form a circle with radius $2L$ 
by identifying $x^-=0$ and $x^-=2L$. 
In order to preserve supersymmetry, we need to impose the same 
boundary condition on scalars $\phi_{ij}$ and spinors $\psi_{ij}$. 
It is in general necessary to choose periodic boundary conditions 
on bosonic field and to retain zero modes, 
if one wishes to take into account the possibility of 
 vacuum condensate or spontaneous symmetry breaking \cite{MaYa}. 
Since we are primarily interested in physical mass spectrum, 
we neglect the zero modes in the present work. 
We shall choose periodic boundary 
conditions for both scalars $\phi_{ij}$ and spinors $\psi_{ij}$, 
leaving the problem of zero modes for a further study 
\begin{equation}
\phi_{ij}(x^{-}) = \phi_{ij}(x^{-}+2L),\quad 
\psi_{ij}(x^{-}) = \psi_{ij}(x^{-}+2L).
\end{equation}
The allowed momenta become discrete and the momentum integral 
is replaced by a summation, 
\begin{equation}
k_{n}^{+} = \frac{\pi}{L} n ,\quad n=1,2,3,....\:, \qquad 
\int_{0}^{\infty} dk^{+} \rightarrow 
 \frac{\pi}{L}\sum_{n =1}^{\infty}. 
\label{momentumsum}
\end{equation}
Then mode expansions (\ref{eq:modephi}) and (\ref{eq:modepsi}) 
for $\phi_{ij}$ and $\psi_{ij}$ become discretized 
\begin{equation}
\phi_{ij} =  \frac{1}{\sqrt{4\pi}} \sum_{n=1}^{\infty}
\frac{1}{\sqrt{n}} \bigg[ A_{ij}(n){\rm e}^{-i\pi n x^{-}/L} +
A_{ji}^{\dag}(n) {\rm e}^{i\pi nx^{-}/L}\bigg],
\end{equation}
\begin{equation}
\psi_{ij} =  \frac{1}{\sqrt{4L}}
\sum_{n=1}^{\infty} \bigg[ B_{ij}(n){\rm e}^{-i\pi nx^{-}/L} +
B_{ji}^{\dag}(n) {\rm e}^{i\pi nx^{-}/L}\bigg], 
\end{equation}
\begin{equation}
A_{ij}(n) = \sqrt{\pi /L} a_{ij}(k^{+}=\pi n/L),\quad 
B_{ij}(n) = \sqrt{\pi /L} b_{ij}(k^{+}=\pi n/L), 
\label{discreteoscillator}
\end{equation}
\begin{equation}
\left[A_{ij}(n),A_{lk}^{\dag}(n')\right] =
\left\{B_{ij}(n),B_{lk}^{\dag}(n')\right\} =
\delta_{nn'}\delta_{il}\delta_{jk} .
\end{equation}  

Let us define the supercharge in this discretized light-cone 
quantization. 
The first supercharge $Q_1$  
in this compactified space is given by 
\begin{equation}
Q_1 =
 2^{1/4}i \sqrt{\pi \over L}\sum_{n=1}^{\infty} \sqrt{n} 
\left[A_{ij}(n)B_{ij}^{\dag}(n)
-A_{ij}^{\dag}(n)B_{ij}(n)\right] .
\label{discretesuperchargeone}
\end{equation}
Since the elimination of gauge field $A_+$ introduces a singular 
factor $1/\partial_-$ in supercharge $Q_2$, 
we need to specify 
an infrared regularization for this factor. 
Following the procedure of 'tHooft \cite{THooft}, we 
employ the principal value prescription for the supercharge. 
Namely we simply drop the zero momentum mode 
\begin{eqnarray}
Q_2 &\!\!\!
=&\!\!\!
 2^{1/4}(-i)g\sqrt{L \over \pi}\sum_{m=1}^{\infty} \frac{1}{m} 
\left[B_{ij}^{\dagger}(m) \tilde J_{ij}(-m)
-\left(\tilde J_{ij}(-m)\right)^{\dagger}B_{ij}(m)\right]\nonumber \\ 
&\!\!\!
=
&\!\!\!
-i\frac{2^{-1/4}g}{\pi}\sqrt{L}\Bigg(
\sum_{l,n=1}^{\infty}\frac{l+2n}{2l\sqrt{n(l+n)}}  \biggl[
\left(A^{\dagger}(n)B^{\dagger}(l)
-B^{\dagger}(l)A^{\dagger}(n)\right)_{ij}A_{ij}(l+n) \nonumber \\ 
&\!\!\!
&\!\!\!
\;\;\;\;\;\;\;\;\;\;\;\;\;\;\;\;\;\; 
-A_{ij}^{\dagger}(l+n)\Bigl(A(n)B(l)
-B(l)A(n)\Bigr)_{ij}\biggr]\nonumber \\
&\!\!\!
+
&\!\!\!
\sum_{l=3}^{\infty}\sum_{n=1}^{l-1}\frac{l-2n}{2l\sqrt{n(l-n)}}
\left[B_{ij}^{\dagger}(l)\biggl(A(n)A(l-n)\biggr)_{ij}
-\Bigl(A^{\dagger}(n)A^{\dagger}(l-n)\Bigr)_{ij}B_{ij}(l)
\right]\nonumber \\
&\!\!\!
-
&\!\!\!
\sum_{l,n=1}^{\infty}\left(\frac{1}{l}
+\frac{1}{n}\right)\biggl[\left(
B^{\dagger}(n)B^{\dagger}(l)\right)_{ij}B_{ij}(l+n)
+B_{ij}^{\dagger}(l+n)\Bigl(B(n)B(l)\Bigr)_{ij}\biggr]\nonumber \\
&\!\!\!
+
&\!\!\!
\sum_{l=2}^{\infty}\sum_{n=1}^{l-1}\frac{1}{l}\left[
B_{ij}^{\dagger}(l)\Bigl(B(n)B(l-n)\Bigr)_{ij}
+\left(B^{\dagger}(n)B^{\dagger}(l-n)\right)_{ij}B_{ij}(l)\right]\Bigg).
\label{discretesuperchargetwo}
\end{eqnarray}

The supersymmetry algebra requires a relation between 
supercharges and the light-cone momentum $P^+$ 
and the Hamiltonian $P^-$ operators 
\begin{equation}
\{Q_1, Q_1\}=2\sqrt2 P^+ ,
\quad 
\{Q_2, Q_2\}=2\sqrt2 P^- ,
\quad 
\{Q_1, Q_2\}=0 ,
\end{equation}
in our choice of spinor notations (\ref{gammamatrix}). 
Infrared regularizations of $P^+$ and $P^-$ have to be done 
consistently with the supersymmetry algebra. 
It is actually difficult to guess the correct infrared regularization 
for the Hamiltonian unless we start from the supercharge. 
The Hamiltonian $P^-$ can be defined by just squaring the supercharge 
$Q_2$. 
Then the above principal value prescription for the supercharge
 $Q_2$ specifies uniquely the prescription for the Hamiltonian. 
In this way we can check that the supersymmetry algebra holds 
in our formulation of the discretized light-cone quantization. 

Physical states take the following form
\begin{equation}
\frac{1}{N^{m/2}\sqrt{s}}\, {\rm tr}\left[{\cal O}(n_{1}) \cdots{\cal O}(n_{m})
\right]\left|0\right\rangle,\:\:\:\:\   m > 1, 
\label{state}
\end{equation}
where $\cal O$ represents $A^{\dag}$ or $B^{\dag}$. 
The symmetry factor $s$ is the number of possible permutations 
of constituents which give the same state \cite{DaKl}. 
Note that we should consider only states with two or more constituents 
$m > 1$ since we should discard singlet to the leading order of 
the $1/N$ expansion of $U(N)$ gauge theory. 
It is also absent in the case of $SU(N)$ gauge theory anyway. 
All these states satisfy the physical 
state condition coming from the constraint (\ref{shiki2})
Here we note that there are both bosonic and 
fermionic oscillators  in our supersymmetric theory. 
This fact gives rise to much larger number of new physical states 
compared to the purely fermionic or bosonic adjoint matter 
case. 

%
%
%
\twelverm   
\baselineskip=14pt
\section{
Numerical Results of Supercharge Diagonalization
}
\vglue 1pt
As we have seen, our procedure preserves supersymmetry manifestly 
throughout the calculation. 
Therefore we are naturally led to obtain supersymmetric mass 
spectra with exactly the same bosonic and fermionic spectra for 
color singlet states. 

If we consider the states with finite values of the discrete momentum 
$K$, we have only finitely many physical states to diagonalize 
the mass matrix.

We have explicitly constructed bosonic and fermionic color singlet 
states for higher values of the cut-off momentum $K$ up to $K=11$. 
We find the number of bosonic color singlet states for 
$K=5, 6, 7, 8, 9, 10,$ and $ 11$ to be $24$, $61$, $156$, $409$, $1096$, 
$2953$, and $ 8052$ respectively. 
The number of fermionic color singlet states is exactly the same 
as the corresponding bosonic one with the same $K$. 

\begin{figure}
 \caption{
The accumulated number of bound states as a function of 
mass 
for $K=4, 5, 6, 7, 8$; there is no differnce in behavior between bosonic
and fermionic state. 
}
 \leavevmode
 \epsfysize=18cm
 \centerline{\epsfbox{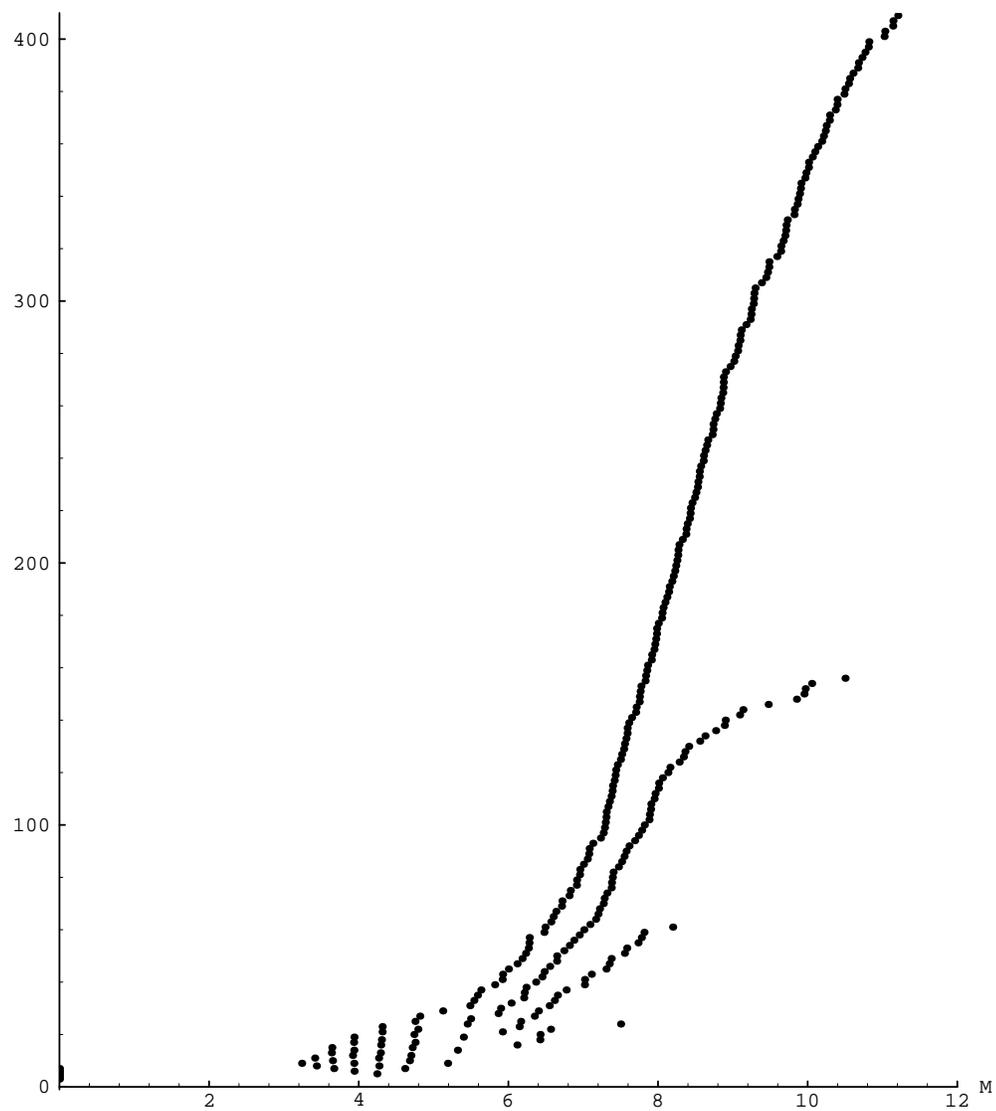}}
 \label{rui}
\end{figure}

After evaluating the supercharge for these subspace up to $K=8$, 
we diagonalize the supercharge exactly to obtain the mass 
eigenvalues. 
In Fig.\ref{rui} we plot the accumulated number of bosonic color 
singlet bound states as a function of mass divided by 
$g\sqrt{N \over \pi}$. 

We can see that the number of states is approaching to a limiting value 
at least for smaller values of $M^2$. 
The present tendency seems to suggest that the density of states is 
increasing rapidly as the mass squared increases. 
This behavior is in qualitative agreement with 
the previous results 
for the adjoint scalar or adjoint spinor matter constituents in 
nonsupersymmetric gauge theories \cite{DaKl}. 
Namely the density of states showed an exponential increase 
as mass squared increases in accordance with the closed string 
interpretation. 
From this result we find numerically that the limiting (Hagedron) 
temperature for the string 
is given by 
\begin{equation}
\beta_H=0.676 \sqrt{\pi \over g^2 N}
\end{equation}
The fermionic color singlet bound states show the same behavior. 

%
%

%
\twelverm   
\baselineskip=14pt
\section{
Weyl Gauge and Fundamental Domain
}
\vglue 1pt
In this section we compactify the spatial 
direction to a circle with a finite radius $L/2\pi$. 
The gauge fields naturally follow periodic 
boundary conditions 
\begin{equation}
A^\mu (x=0) = A^\mu (x=L) .
\end{equation}
We shall specify boundary conditions for $\Psi$ and $\phi$ later. 
%

Gauge theories have a large number of redundant 
gauge degrees of freedom which should be eliminated by a gauge-fixing 
condition.
In this section we quantize the system in the Weyl gauge,
\begin{equation}
A^0 = 0 .
\end{equation}
We can impose Gauss' law as a subsidiary condition for the 
physical state $| \Phi \rangle$ 
\begin{equation}
[ D_1 E^a(x) -g \rho^a(x) ] | \Phi \rangle =0, 
\quad 
\rho^a 
=
 f^{abc} \phi^b \pi^c
+ \frac{i}{2} f^{abc} \Psi_\alpha^c \Psi_\alpha^b.
\label{Gauss cond.} 
\end{equation}
where $\pi^a$ and $-E^a\equiv F^{a01}$ are the conjugate variables 
of $\phi^a$ and $A^{a1}$ respectively, and 
$\rho^a$ is the color charge density, 
and $f^{abc}$ are the structure constants of the Lie algebra of $SU(N)$ : 
 $[t^a, t^b] = i f^{abc} t^c$. 
Note that the Gauss law determines $E$, except for 
its constant modes $e$. 
One can eliminate $A^1$ by using an appropriate gauge transformation, 
except for the $N-1$ spatially constant modes $a^p$ which are given by 
\begin{equation}
{\cal P}\exp \left( i g \int_0^L dx A^1(x) \right) = 
V e^{i g a L} V^\dagger ,\quad (a = a^p t^p),\label{a^p def}
\label{constantmode;a}
\end{equation}
where $V$ is a unitary matrix. 
Hereafter we shall use the convention that 
$a, b, \cdots =1,2 \cdots, N^2-1$ 
represent the indices of the generators of $SU(N)$, 
and $p,q, \cdots =1,2 \cdots, N-1$ represent those of Cartan subalgebra. 
The commutation relation between $a^p$ and 
$e^q$ is given as \cite{LeNaTh} 
\begin{equation}
\left[ e^p , a^q \right] = i \delta^{p q} \qquad
p,q = 1,\ldots,N-1.
\end{equation}

In the physical state space, we can eliminate 
redundant gauge degrees of freedom by solving 
the Gauss law constraint (\ref{Gauss cond.}), and find 
the Hamiltonian 
\begin{eqnarray}
H
&\!\!\!
=
&\!\!\!
\int_0^L dx {\cal H}(x) 
=
K_a + H_{\rm c} + H_{\rm b} + H_{\rm f} + H_{\rm int} , 
\label{hamiltonian;1} \\
K_a 
&\!\!\!
=
&\!\!\!
 \frac{1}{2L} \sum_p e^{p \dagger} e^p,\\
H_{\rm c} &=& \frac{g^2}{L} \sum_{n=- \infty}^{\infty}
\sum_{ij} \int_0^L dy \int_0^L dz 
(1 - \delta_{ij} \delta_{n0} )
\frac{ \left( \rho(y) \right)_{ij} \left( \rho(z) \right)_{ji}}
{\left( \frac{2 \pi n}{L} + g (a_i - a_j) \right)^2}
e^{2 \pi i n (y-z)/L} ,\label{H_c} \\
H_{\rm b} 
&\!\!\!
=
&\!\!\!
 \int_0^L dx \left\{ 
\frac{1}{2} \pi^a \pi^a + 
\frac{1}{2} \left(D_1 \phi \right)^a 
\left( D_1 \phi \right)^a \right\},\\
H_{\rm f} 
&\!\!\!
=
&\!\!\!
 \int_0^L dx \left(- \frac{i}{2} \right)
\Psi^a \sigma_3 \left( D_1 \Psi \right)^a , \qquad
H_{\rm int} 
=
 \int_0^L dx \,
{\rm tr} \left\{ i g \phi \bar{\Psi} \gamma_5 \Psi \right\} 
\label{H_int}
\end{eqnarray}
where $ a_i = a^p t^p_{ii}$  with no summation over $i$ implied, 
and $\sum_i a_i = 0$. 
Here the covariant derivative $D_1$ contains only the zero mode of 
$A^1$ : $D_1 = \partial_1 - i g [a,~]$. 
One should note that gauge fields $A^\mu$, except the zero modes $a^p$, 
are completely eliminated.

In order to investigate the vacuum structures of our model, we solve 
Schr\"odinger's equation with respect to the Hamiltonian 
(\ref{hamiltonian;1}) 
\begin{eqnarray}
H| \Phi \rangle = E |\Phi \rangle ,
\end{eqnarray}
where $|\Phi \rangle$ denote state vectors in the physical space.
Because of hermiticity of the variables $a$, 
the kinetic energy $K_a$ is given in terms of the 
Jacobian $J[a]$ of the transformation (\ref{constantmode;a}) 
\cite{LeNaTh} 
\begin{eqnarray}
K_a 
&\!\!\!
= 
&\!\!\!
\frac{1}{2L} e^{p \dagger} e^p 
= - \frac{1}{2L} \frac{1}{J[a]} \frac{\partial}{\partial a^p}
J[a] \frac{\partial}{\partial a^p}, \\
J[a] 
&\!\!\!
=
&\!\!\!
 \prod_{i>j} \sin^2 
\left( \frac{1}{2} g L (a_i - a_j) \right). 
\label{J}
\end{eqnarray}

In analogy with the radial wavefunctions, it is useful to define 
a modified wave function 
\begin{equation}
\tilde{\Phi}[a] \equiv \sqrt{J[a]}\Phi [a] .
\label{radial}
\end{equation}
The kinetic energy operator for $\tilde \Phi$ is 
(with the notation $\partial_{p} = \partial/\partial a^{p}$), 
\begin{equation}
K_a^\prime \equiv 
\sqrt{J} K_a {1 \over \sqrt{J}} 
= -\frac{1}{2L} \partial_p \partial_p + V^{[N]} ,
\label{K_a^'}
\end{equation}
\begin{equation}
V^{[N]} 
\equiv 
\frac{1}{2L} \frac{1}{\sqrt{J}}\left(\partial_p \partial_p \sqrt{J}
\right)
= - \frac{(gL)^{2}}{48L} N (N^{2}-1) .
\label{V}
\end{equation}
Thus we obtain a boundary condition
for the modified wavefunction,
\begin{eqnarray}
\tilde{\Phi}[a]=0, \, \ \ {\rm if} \quad J[a]=0 \ .
\label{boundary}
\end{eqnarray}


Let us now quantize the fields $\Psi$ and $\phi$. 
The gauge field zero modes $a^p$ couple only to 
off-diagonal elements, which are parameterized as : 
$\varphi_{ij} = \sqrt{2} \Psi_{ij}$, 
$\varphi_{ij}^\dagger = \sqrt{2} \Psi_{ji}$, 
$\xi_{ij} = \sqrt{2} \phi_{ij}$, 
$\xi_{ij}^\dagger = \sqrt{2} \phi_{ji}$, 
$\eta_{ij} = \sqrt{2} \pi_{ij}$, and 
$\eta_{ij}^\dagger = \sqrt{2} \pi_{ji}$
$(i<j)$.
With these conventions the Hamiltonian takes the form
\begin{eqnarray}
H_{\rm f} 
&\!\!\!
=
&\!\!\!
 H_{\rm f,diag} + H_{\rm f,off}, \qquad 
H_{\rm b} = H_{\rm b,diag} + H_{\rm b,off} ,\\
H_{\rm f,diag} 
&\!\!\!
=
&\!\!\!
 \frac{1}{2i} 
\sum_p \int_0^L dx \Psi^p \sigma_3 \partial_1 \Psi^p ,\label{H_f,diag} \\
H_{\rm f,off} 
&\!\!\!
=
&\!\!\!
 \sum_{i<j} \int_0^L dx \varphi_{ij}^\dagger 
\sigma_3 \left( \frac{1}{i} \partial_1 - g(a_i - a_j) \right) 
\varphi_{ij} ,\label{H_f,off} \\
H_{\rm b,diag} 
&\!\!\!
=
&\!\!\!
 \sum_p \int_0^L dx
\left( \frac{1}{2} \pi^p \pi^p + 
\frac{1}{2} \left(\partial_1 \phi^p \right) 
\left( \partial_1 \phi^p \right) \right),\label{H_b,diag} \\
H_{\rm b,off} &=& \sum_{i<j} \int_0^L dx \Bigl\{
\eta_{ij}^\dagger \eta_{ij} 
{}+ \left(\partial_1 \xi_{ij}^\dagger 
- i g (a_j - a_i) \xi_{ij}^\dagger \right)
\left(\partial_1 \xi_{ij}
- i g (a_i - a_j) \xi_{ij} \right) \Bigr\}.\label{H_b,off}
\end{eqnarray}

Let us now discuss the range \cite{Lenz} 
of the variables $a^p$. 
Eq.(\ref{a^p def}) shows that 
the $g L a$ are angular variables
which are defined only in modulo $2 \pi$.
If the parameterization of $a$ is one-to-one 
and permutations of the eigenvalues are contained in a single domain, 
the domain is called {\it the elementary cell}. 
{}For example, in the SU(2) case, 
two eigenvalues of the matrix $a$ are 
$a_1 = a^3/2$ and $a_2 = -a^3/2$. 
Then, the elementary cell is the interval 
$
- \pi \le {gLa^3 \over 2} \le \pi,
$
with the end points identified. 
If $a^3$ is negative in the elementary cell, 
the Weyl reflection $a^3 \to -a^3$ 
maps the interval 
$-{2\pi \over gL}<a^3<0$ onto the interval $0<a^3<{2\pi \over gL}$ 
(simultaneously, 
$\varphi^{12}\leftrightarrow\varphi^{21}$). 
In the $SU(N)$ case, similarly, 
the elementary cell is divided into $N!$ domains by 
the Weyl group since the Weyl group 
of $SU(N)$ is the permutation group $P_N$. 
These $N!$ domains are called {\it fundamental domains}. 
Boundaries of the fundamental domains consist of the hypersurfaces 
where two of the eigenvalues match. 
If two of the eigenvalues have the same value, the Jacobian 
$J[a]$ vanishes. 
In the case of $SU(2)$, we take the following interval 
as the fundamental region 
\begin{eqnarray}
0 \le a^{3} \le \frac{2\pi}{gL} .
\end{eqnarray}
The Jacobian $J[a]={\rm sin}^2 \left({1 \over 2}gLa^3 \right)$ 
vanishes at $a^3=0,~\frac{2\pi}{gL}$. 
Note that the modified wavefunction $\tilde{\Phi}[a]$ vanishes
at these points.

%
\twelverm   
\baselineskip=14pt
\section{
Vacuum Structures of SUSY $SU(2)$ Yang-Mills 
Theories
}
\vglue 1pt
In this section, we determine the wave function of the vacuum state 
in the fundamental domain 
by using the Born-Oppenheimer approximation \cite{Lenz}. 
If $gL \ll 1$, the energy scale of the system of $a^p$ is given 
by $(gL)^2/L$, while that of 
non-zero modes of
$\Psi$ and $\phi$ is in general of order 
$1/L$.
Therefore we can integrate the non-zero modes of $\Psi$ and $\phi$
to obtain the effective potential for $a^p$. 
We will retain the zero modes of $\Psi$ and $\phi$,
since their spectrum is continuous. 
By solving the Schr\"odinger equation with respect to the 
resulting effective potential, 
we obtain the wavefunction $\tilde\Phi[a]$, which describes the 
vacuum structures of our model. 
In these procedures we must pay attention to the boundary conditions 
for $\tilde\Phi[a]$ resulting from the Jacobian 
(\ref{boundary}). 

To calculate the effective potential as a function of 
the gauge zero modes $a^p$, 
we have to find the ground state of fermion $\Psi$ and boson $\phi$ 
for a fixed value of $a^p$. 
Here, we must take care with regards to the boundary conditions 
for $\Psi(x)$ and 
$\phi(x)$. 
Since spinors and scalars are superpartners of gauge fields 
which obey the periodic boundary condition, the spinors $\Psi(x)$ and 
scalars $\phi(x)$ should be periodic in order for the boundary 
conditions to maintain supersymmetry 
\begin{eqnarray}
\Psi(x=0) = \Psi(x=L) , \qquad 
\phi(x=0) = \phi(x=L).
\end{eqnarray}
Hereafter we refer to this boundary condition as {\it the (P,P) case}.
In this section we investigate the vacuum structures
for the  
gauge group $SU(2)$. 

\twelverm   
\baselineskip=14pt
\subsection{
Born-Oppenheimer Approximation 
}
\vglue 1pt
{}For $gL \ll 1$, 
the Coulomb energy (\ref{H_c}) 
and the Yukawa interaction (\ref{H_int}) can be neglected.
In this limit, the relevant parts of the Hamiltonian are, for $SU(2)$, 
\begin{eqnarray}
\tilde{H}
&\!\!\!
 =
&\!\!\!
 K_a^\prime + H_{\rm b,diag} + H_{\rm b,off}
+ H_{\rm f,diag} + H_{\rm f,off} .\label{tilde H} \\
K_a^\prime 
&\!\!\!
=
&\!\!\!
 -\frac{1}{2L}
\frac{\partial^2}{\partial a^2} + V^{[N=2]}, \\
H_{\rm b,diag} 
&\!\!\!
=
&\!\!\!
 \frac{1}{2} \int_0^L dx 
\left\{ \pi^3 \pi^3 +(\partial_1 \phi^3)(\partial_1 \phi^3)
 \right\}\label{b,diag} \\
H_{\rm b,off} 
&\!\!\!
=
&\!\!\!
 \int_0^L dx
\left\{ \eta^\dagger \eta + (\partial_1 \xi^\dagger +iga\xi^\dagger)
(\partial_1 \xi - iga\xi) \right\} ,\label{b,off} \\
H_{\rm f,diag} 
&\!\!\!
=
&\!\!\!
 \frac{1}{2i} \int_0^L dx
\Psi^3 \sigma_3 \partial_1 \Psi^3, \label{f,diag}
\qquad
H_{\rm f,off} 
=
 \int_0^L dx
\varphi^\dagger \sigma_3 \left( \frac{1}{i}
\partial_1
 -ga \right) \varphi, \label{f,off} \\
 \varphi 
&\!\!\!
\equiv
&\!\!\!
 \varphi_{12},\quad
\xi \equiv \xi_{12},\quad
\eta \equiv \eta_{12}, \quad
a \equiv a^3 =a_1 - a_2 . 
\end{eqnarray}

A remnant of large gauge transformations becomes a discrete symmetry 
$S$ \cite{Lenz} 
\begin{eqnarray}
S:
&\!\!\!
&\!\!\!
\quad a \to -a+\frac{2 \pi}{gL} , \nonumber \\
\varphi 
&\!\!\!
\to
&\!\!\!
 e^{2i \pi x/L} \varphi^{\dagger} \ ,
\quad
\xi \to e^{2i \pi x/L} \xi^\dagger \ ,
\quad
\eta \to e^{2i \pi x/L} \eta^\dagger \ ,
\nonumber \\
\Psi^3 
&\!\!\!
\to 
&\!\!\!
-\Psi^3, 
\quad
\phi^3 \to -\phi^3, 
\quad
\pi^3 \to -\pi^3.
\label{S}
\end{eqnarray}
This operator can be chosen to satisfy 
$S^2 =1$ and $[S,H]=0$. 
SYM$_2$ 
has a topologically nontrivial structure $\pi_1[SU(N)/Z_N]=Z_N$. 
The symmetry $S$ corresponds to a nontrivial element of 
this $Z_{N=2}$ group for $SU(2)$. 


In order to perform the Born-Oppenheimer approximation, we first expand 
the spinor fields $\varphi$ and $\Psi^3$, and impose a canonical 
anticommutation relation 
\begin{eqnarray}
\varphi\left(x\right)
&\!\!\!
=
&\!\!\!
\frac{1}{\sqrt{L}}\sum_{k=-\infty}^\infty{a_k \choose b_k}
e^{i 2 \pi k x /L} , 
\qquad
\left\{ a_k,a_{k^\prime}^\dagger \right\} =
\left\{ b_k,b_{k^\prime}^\dagger \right\} =
\delta_{k,k^{\prime}} ,
\nonumber \\
\Psi^3\left(x\right)
&\!\!\!
=
&\!\!\!
\frac{1}{\sqrt{L}}\sum_{k=-\infty}^\infty{c_k \choose d_k}
e^{i 2 \pi k x /L} ,
\qquad c_{-k}=c_k^\dagger ,
\quad d_{-k}=d_k^\dagger ,
\label{expand;fermion} \\
&\!\!\!
&\!\!\!
\left\{ c_k,c_{k^\prime}^\dagger \right\} =
\left\{ d_k,d_{k^\prime}^\dagger \right\} =
\delta_{k,k^{\prime}},\quad
k, k^{\prime} \ge 0 \nonumber
\end{eqnarray}
The Hamiltonian $H_{\rm f,off}$ in (\ref{f,off}) takes the form 
\begin{eqnarray}
H_{\rm f,off} = \sum_{k =-\infty}^{\infty}
\left(a_k^\dagger a_k - b_k^\dagger b_k \right)
\left(\frac{2 \pi k}{L} - ga \right) .
\end{eqnarray}
In the Born-Oppenheimer approximation,
the vacuum state for the off-diagonal part of the fermion
is obtained by filling the Dirac sea for the fermion $\varphi$. 
We assume the $a_k$ modes to be filled for $k<M$. 
The Gauss law constraint (\ref{Gauss cond.}) dictates that the 
$b_k$ modes should be filled for $k \ge M$ \cite{Lenz}.
Denoting the vacuum state for the fermion as 
$| 0_\varphi ;M \rangle$, the vacuum energy can be written as 
\begin{eqnarray}
H_{\rm f,off} | 0_\varphi ;M \rangle &=& \left[
\sum_{k=-\infty}^{M-1} \left(\frac{2 \pi k}{L} -ga \right)
- \sum_{k=M}^{\infty} \left(\frac{2 \pi k}{L} -ga \right)
\right] | 0_\varphi ;M \rangle \nonumber \\
&\equiv& V_{\rm f,off}(a;M)| 0_\varphi ;M \rangle .
\label{ev of f,off}
\end{eqnarray}
Notice that $S$ acts on the state $| 0_\varphi ;M \rangle$ 
according to 
\begin{eqnarray}
S |0_\varphi ;M \rangle
= e^{i \alpha_M} | 0_\varphi ; 2-M \rangle .
\label{actionS;pp}
\end{eqnarray}
In addition, the phase factor $e^{i \alpha_M}$ is constrained 
by $S^2=1$, or in other words, $e^{i \alpha_M} = e^{-i \alpha_{-M+2}}$. 

{}For diagonal part of the fermion, 
we obtain the Hamiltonian from (\ref{f,diag}) 
\begin{eqnarray}
H_{\rm f,diag} = \sum_{k \ge 1} \frac{2 \pi k}{L}
\left( c_k^\dagger c_k + d_k d_k^\dagger -1 \right) .
\end{eqnarray}
On the vacuum $|0_{\Psi}\rangle$ defined by 
$c_k|0_{\Psi}\rangle=d^\dagger_k|0_{\Psi}\rangle=0,~~ k\ge 1$,
we find
\begin{eqnarray}
H_{\rm f,diag}|0_{\Psi} \rangle = - \sum_{k \ge 1} \frac{2 \pi k}{L}
|0_{\Psi} \rangle 
\equiv V_{\rm f,diag} |0_{\Psi} \rangle.
\label{ev of f,diag}
\end{eqnarray}


Next we expand the scalar fields $\xi$, $\eta$, $\phi^3$, and $\pi^3$, 
and impose canonical commutation relations 
\begin{eqnarray}
\xi\left(x\right) 
&\!\!\!
=
&\!\!\!
\sum_{k=-\infty}^\infty \frac{1}{\sqrt{2L E_k}}
\left(e_k + f_k^\dagger \right)
e^{i 2 \pi k x /L} ,
\quad E_k=\left|\frac{2 \pi k}{L} -ga \right| ,\\
\eta\left(x\right) 
&\!\!\!
=
&\!\!\!
\sum_{k=-\infty}^\infty i \sqrt{\frac{E_k}{2L}}
\left(- e_k + f_k^\dagger \right)
e^{i 2 \pi k x /L} ,\\
\phi^3\left(x\right) 
&\!\!\!
=
&\!\!\!
\sum_{k=-\infty \atop k \ne 0}^\infty \frac{1}{\sqrt{2L F_k}}
\left(g_k + g_{-k}^\dagger \right)
e^{i 2 \pi k x /L} 
+ \phi_{\rm zero} ,
\quad F_k=\left|\frac{2 \pi k}{L} \right| ,\\
\pi^3\left(x\right) 
&\!\!\!
=
&\!\!\!
\sum_{k=-\infty \atop k \ne 0}^\infty i \sqrt{\frac{F_k}{2L}}
\left(- g_k + g_{-k}^\dagger \right)
e^{i 2 \pi k x /L} 
+\frac{1}{L} \pi_{\phi_{\rm zero}} .
\end{eqnarray}
\begin{eqnarray}
\left[ e_k,e_{k^\prime}^\dagger \right] =
\left[ f_k,f_{k^\prime}^\dagger \right] =
\left[ g_k,g_{k^\prime}^\dagger \right] =
\delta_{k,k^{\prime}}, \quad 
\left[ \phi_{\rm zero} , \pi_{\phi_{\rm zero}} \right] = i .
\end{eqnarray}

The Hamiltonian $H_{\rm b,off}$ in (\ref{b,off}) is given by 
\begin{eqnarray}
H_{\rm b,off}
&\!\!\!
=
&\!\!\!
 \sum_{k = -\infty}^{\infty}
E_k \left( e_k^\dagger e_k + f_k f_k^\dagger \right) \\
&\!\!\!
=
&\!\!\!
 \sum_{k = -\infty}^{\infty}
E_k \left( e_k^\dagger e_k + f_k^\dagger f_k \right) 
{}- \sum_{k=-\infty}^{N-1} \left( \frac{2 \pi k}{L} - ga \right)
+ \sum_{k=N}^{\infty} \left( \frac{2 \pi k}{L} -ga \right) ,
\end{eqnarray}
where $N$ is an integer satisfying 
\begin{equation}
\frac{2 \pi N}{L} -ga \ge 0 ,\qquad \frac{2 \pi (N-1)}{L} -ga < 0 .
\label{N}
\end{equation}
On the vacuum state $|0_{\xi}\rangle$ defined by 
$e_k |0_\xi \rangle = f_k |0_\xi \rangle = 0,~~ \mbox{for all }k$, 
we find the vacuum energy 
\begin{eqnarray}
H_{\rm b,off} | 0_\xi \rangle &=& \left[
- \sum_{k=-\infty}^{N-1} \left( \frac{2 \pi k}{L} - ga \right)
+ \sum_{k=N}^{\infty} \left( \frac{2 \pi k}{L} -ga \right)
\right] | 0_\xi \rangle \nonumber \\
&\equiv& V_{\rm b,off}(a)| 0_\xi \rangle .
\label{ev of b,off}
\end{eqnarray}

We find that the zero mode Hamiltonian $H_0$ is separated as 
\begin{eqnarray}
H_{\rm b,diag} 
&\!\!\!
= 
&\!\!\!
\sum_{k \ge 1} \frac{2 \pi k}{L}
\left( g_k^\dagger g_k + g_{-k}^\dagger g_{-k} +1 \right) +H_0 , \\
H_0 
&\!\!\!
=
&\!\!\!
 \frac{1}{2L} \pi_{\phi_{\rm zero}} \pi_{\phi_{\rm zero}} .
\label{zeromodehamiltonian}
\end{eqnarray}
On the vacuum for the nonzero modes of $\phi^3$ 
satisfying $g_k|0_{\phi} \rangle = g_{-k}|0_{\phi}\rangle =0 , k\ge 1$, 
we find the vacuum energy
\begin{eqnarray}
V_{\rm b,diag} = \sum_{k \ge 1} \frac{2 \pi k}{L}.
\end{eqnarray}

\twelverm   
\baselineskip=14pt
\subsection{
Vacuum Structure 
}
\vglue 1pt
The vacuum energies obtained in the previous section 
are divergent. 
By regularizing them with
the heat kernel, 
we obtain the following finite 
effective potential as a function of $a$ 
\begin{eqnarray}
U_{M,N}(a)
&\!\!\!
=
&\!\!\!
V_{\rm f,off}(a;M) + V_{\rm b,off}(a) +
V_{\rm f,diag} + V_{\rm b,diag} 
+V^{[N=2]} 
\nonumber \\
&\!\!\!
=
&\!\!\!
\frac{2 \pi}{L} 
\left( M - \frac{gLa}{2 \pi} - \frac{1}{2} \right)^2
-\frac{2 \pi}{L} 
\left( N - \frac{gLa}{2 \pi} - \frac{1}{2} \right)^2
+V^{[N=2]} .
\label{effpotzeromode}
\end{eqnarray}
In the fundamental region $0<{gLa \over 2}< \pi$, $N=1$ 
from (\ref{N}). 
By requiring that the vacuum energy $U_{M,N}(a)$ 
be minimal,
we can fix $M$ to obtain $M=1$. 
We then find that the total vacuum energy in the 
fundamental domain is independent of $a$ 
\begin{equation}
U_{M,N}(a)=
V^{[N=2]} .
\end{equation}
Consequently we obtain the Hamiltonian which describes the vacuum 
structures for the periodic boundary condition 
\begin{equation}
\tilde{H} = K_a^\prime + H_0 
=-\frac{1}{2L}\frac{\partial}{\partial a}
\frac{\partial}{\partial a} +V^{[N=2]} +
\frac{1}{2L} \pi_{\phi_{\rm zero}} \pi_{\phi_{\rm zero}} .
\label{zeroH}
\end{equation}
We also have the zero modes of the fermion, which form a Clifford 
algebra 
\begin{eqnarray}
\Psi_{\rm zero}^3
&\!\!\!
=
&\!\!\!
 \frac{1}{\sqrt{L}}{c_0 \choose d_0},
\qquad c_0=c_0^\dagger ,
\quad d_0=d_0^\dagger ,\\
\left\{ \lambda , \lambda^\dagger \right\} 
&\!\!\!
=
&\!\!\!
 1 ,\quad
\left\{ \lambda , \lambda \right\}=
\left\{ \lambda^\dagger , \lambda^\dagger \right\}=0 ,\quad 
\lambda \equiv \frac{1}{\sqrt{2}} ( c_0 + i d_0 ) . 
\label{clifford;algebra}
\end{eqnarray}
Let us now solve the Schr\"odinger equation 
\begin{equation}
\tilde{H}\tilde{\Phi}(a)=e\tilde{\Phi}(a).
\end{equation}
Because of the boundary condition (\ref{boundary}) 
we get the wavefunction $\tilde \Phi(a)$ and the energy eigenvalue $e$ 
of the ground state as 
\begin{equation}
\tilde{\Phi}(a) = \sqrt{\frac{gL}{\pi}} 
\sin \left( \frac{gLa}{2} \right),
\qquad 
e=0.
\label{the ground state and e}
\end{equation}
It is interesting to note that the vacuum energy 
associated with the nontrivial zero mode wavefunction
(\ref{the ground state and e})
cancels precisely the contribution $V^{[N=2]}$ from the 
Jacobian in (\ref{V}).
Therefore we have shown explicitly that the SUSY is not broken 
spontaneously. 
Also note that our result is consistent with the previous 
calculation of the nonvanishing Witten index \cite{Li}. 
The calculation, however, ignores the Jacobian (\ref{J}), which is 
an important ingredient in our present attempt to define 
the gauge field zero modes properly \cite{Lenz}. 
Therefore the above explicit demonstration of the vanishing vacuum
energy using the Born-Oppenheimer approximation can be regarded as
another independent proof of 
the unbroken SUSY in SUSY 
Yang-Mills theories in $1+1$ dimensions. 


We define the vacuum 
state of the zero modes of the fermion $c_0, d_0$. Note that the zero 
modes belong to the two-dimensional representation of the Clifford 
algebra (\ref{clifford;algebra}). 
We define $| \Omega \rangle$ to be the Clifford vacuum 
annihilated by $\lambda$ and $| \tilde{\Omega} \rangle =
\lambda^\dagger | \Omega \rangle$.
Since the field $\phi^3$ can take unbounded values,
the zero mode spectrum is continuous.
This fact makes the Witten index ill-defined.
The previous attempt to compute the Witten index employed
a regularization by putting a cut-off on the $\phi_{\rm zero}$ space.
In that case, the Witten index can be defined and obtains
${\rm tr}(-1)^{F} = 1$ \cite{Li}.
In spite of this complication, we can choose the wave function
to be constant in the $\phi^3$ zero mode as the vacuum:
$H_0 | \omega \rangle =0$.

Let us now examine the transformation property under the 
discrete gauge transformation $S$.
The non-zero mode vacuum $|0_\varphi; M=1 \rangle$
turns out to be an eigenstate of $S$
\begin{equation}
S|0_\varphi; M=1 \rangle = \pm|0_\varphi; M=1 \rangle
\end{equation}
because of eq.(\ref{actionS;pp}) and $S^2=1$.
Similarly $|0_\Psi \rangle$, $|0_\xi \rangle$,
$|0_\phi \rangle$ and $| \omega \rangle$
are eigenstate of $S$ with eigenvalues $\pm 1$.
{}For the fermion zero mode, $S| \Omega \rangle = \pm |\Omega \rangle$
and  $S| \tilde{\Omega} \rangle = \mp | \tilde{\Omega} \rangle$.
Since we should construct the full vacuum state as an eigenstate with
eigenvalue
$\pm 1$ for $S$
\begin{eqnarray}
|{\bf 0}_{\Omega} \rangle 
&\!\!\! \equiv &\!\!\! 
|\tilde{\Phi}(a) \rangle
|0_\varphi ;M=1 \rangle
|0_\Psi \rangle |0_\xi \rangle |0_\phi \rangle
| \omega \rangle
| \Omega \rangle , \nonumber \\
|{\bf 0}_{\tilde{\Omega}} \rangle 
&\!\!\! \equiv &\!\!\! 
|\tilde{\Phi}(a) \rangle
|0_\varphi ;M=1 \rangle
|0_\Psi \rangle |0_\xi \rangle |0_\phi \rangle
| \omega \rangle
| \tilde{\Omega} \rangle .
\end{eqnarray} 
We find the vacuum condensate
$\left| \langle {\bf 0} |  \bar{\Psi}^a \Psi^a 
| {\bf 0}  \rangle \right|= \frac{1}{L}$ for
both $| {\bf 0} \rangle = |{\bf 0}_{\Omega} \rangle$ and
$|{\bf 0}_{\tilde{\Omega}} \rangle$.
One can see that this condensate is due to the finite spacial extent $L$.

\vspace{5mm} 
%
%

We wish to acknowledge Simon Dalley for a useful discussion and  
advise on diagonalization of matrices, Kenichiro Aoki for 
an illuminating discussion and Des Johnston for a reading of 
the manuscript. 
One of the authors (N.S) thanks Tohru Eguchi, Kiyoshi Higashijima, 
Sung-Kil Yang, and Elcio and Christina Abdalla for an interesting 
discussion. 
One of the authors (N.S.) would like to thank the Aspen Center for 
Physics and Service de Physique Theorique Saclay for hospitality, 
and the Japan Society for the Promotion of Science for a grant. 
This work is supported in part by Grant-in-Aid for 
Scientific Research (No.05640334), and 
Grant-in-Aid for Scientific Research for Priority Areas 
(No.05230019) from the Ministry of Education, Science 
and Culture.
\vspace{5mm} 
%
%
%
%

%
%

%


\begin{thebibliography}{9}
%
\bibitem{Nilles} H.P. Nilles, {\it Phys.\ Lett.\ }{\bf B115} (1982) 193; 
   S. Ferrara, L. Girardello and H.P. Nilles, {\it Phys.\ Lett.\ }
{\bf 125B} (1983) 457.
\bibitem{VeYa} G. Veneziano and S. Yankielowicz, {\it Phys.\ Lett.\ }
{\bf B113} (1982) 231; 
   J. Affeleck, M. Dine and N. Seiberg, {\it Nucl.\ Phys.\ }{\bf B241}
 (1984) 493; 
   T.R. Taylor, {\it Phys.\ Lett.\ }{\bf B164} (1985) 43. 
\bibitem{SeWi} N. Seiberg and E. Witten, {\it Nucl.\ Phys.\ }{\bf B426} 
(1994) 19,
               ibid. {\it Nucl.\ Phys.\ }{\bf B431} (1994) 484; 
               C. Vafa and E. Witten, {\it Nucl.\ Phys.\ }{\bf B431} 
(1994) 3. 
\bibitem{THooft} G. 't Hooft, {\it Nucl.\ Phys.\ }{\bf B72} (1974) 461, 
                 ibid. {\it Nucl.\ Phys.\ }{\bf B74} (1974) 461. 
\bibitem{CaCoGr} C.G. Callan, N. Coote and D.J. Gross, 
                 {\it Phys.\ Rev.\ }{\bf D13} (1976) 1649; 
               M.B. Einhorn, {\it Phys.\ Rev.\ }{\bf D14} (1976) 3451. 
\bibitem{Ferrara} S. Ferrara, Lett. Nouvo Cimento {\bf 13} (1975) 629.
\bibitem{DaKl} S. Dalley and I.R. Klebanov, {\it Phys.\ Rev.\ }
{\bf D47} (1993) 2517; 
       G. Bhanot, K. Demeterfi, and I.R. Klebanov, {\it Phys.\ Rev.\ }
{\bf D48} (1993) 4980;
       K. Demeterfi, I.R. Klebanov, and G. Bhanot, {\it Nucl.\ Phys.\ }
{\bf B418} (1994) 15.
\bibitem{PuBr} H-C. Pauli and S. Brodsky, {\it Phys.\ Rev.\ }
{\bf D32} (1985) 1993, 
        ibid 2001; K. Hornbostel, Ph.D Thesis, SLAC report 333 (1988). 
\bibitem{HoBr} 
      K. Hornbostel, S.J. Brodsky, and H-C. Pauli, {\it Phys.\ Rev.\ }
{\bf D41} (1990) 3814; 
      R.J. Perry, A. Harindranath and K.G. Wilson, 
{\it Phys.\ Rev.\ Lett.\ }{\bf 65} (1990) 2959; 
      K.G. Wilson et.al., {\it Phys.\ Rev.\ }{\bf D49} (1994) 6720. 
%
%
\bibitem{LeShTh} A. V. Smilga, {\it Phys.\ Rev.\ }{\bf D49} (1994) 6836;
                 F. Lenz, M. Shifman, and M. Thies, hep-th.9412113. 
\bibitem{MaYa} T. Maskawa and K. Yamawaki, {\it Prog.\ Theor.\ Phys.\ }
{\bf 56} (1976) 270; 
        J.E. Hetrick and Y. Hosotani, {\it Phys.\ Lett.\ }{\bf B230} 
(1989) 88; 
      F. Lenz, M. Thies, S. Levit, and K. Yazaki, {\it Ann.\ Phys.\ }
{\bf 208} (1991) 1;
G. McCartor, {\it Z.\ Phys.\ }{\bf C52} (1991) 611;
      T. Heinzl, S. Krusche, and E. Werner, {\it Phys.\ Lett.\ }
{\bf B272} (1991) 54;
   G. McCartor and D.G. Robertson, {\it Z.\ Phys.\ }{\bf C53} (1992) 679;
      D.G. Robertson,{\it Phys.\ Rev.\ }{\bf D47} (1993) 2549;
      C. Bender, S. Pinsky and B. van de Sanda, {\it Phys.\ Rev.\ }
{\bf D48} (1993) 816, 
      {\it Phys.\ Rev.\ }{\bf D49} (1994) 2001, 
      {\it Phys.\ Rev.\ }{\bf D51} (1995) 726; 
      K. Harada, T. Sugihara, M. Taniguchi, and M. Yahiro, 
{\it Phys.\ Rev.\ }{\bf D49} (1994) 4226;
      A.C. Kalloniatis and D.G. Robertson, {\it Phys.\ Rev.\ }
{\bf D50} (1994) 5262; 
      A.C. Kalloniatis, H-C. Pauli, and S. Pinsky, {\it Phys.\ Rev.\ }
{\bf D50} (1994) 6633;
      M. Maeno, {\it Phys.\ Lett.\ }{\bf B320} (1994) 83; 
        Yoonbai Kim, S. Tsujimaru and K. Yamawaki, preprint DPNU-94-56; 
      M. Tachibana, hep-th.9504026.
%
\bibitem{KoSaSa} S. Kojima, N. Sakai and T. Sakai, 
{\it Prog.\ Theor.\ Phys.\ }{\bf 95} (1996) 621.
%
%
\bibitem{Lenz} F. Lenz, M. Shifman, and M. Thies, {\it Phys.\ Rev.\ }
{\bf D51} (1995) 7060.
%
\bibitem{Smilga} A. V. Smilga, {\it Phys.\ Rev.\ }{\bf D49} (1994) 6836.
%
\bibitem{Wit} E. Witten, {\it Nucl.\ Phys,\ }{\bf B202} (1982) 253.
%
\bibitem{Sm} A. V. Smilga, {\it Sov.\ Phys.\ JETP\ }{\bf 64} (1986) 8;
B. Yu. Blok and A. V. Smilga {\it Nucl.\ Phys.\ }{\bf B287} (1987) 589;
A. V. Smilga {\it Nucl.\ Phys.\ }{\bf B291} (1987) 241. 
%
\bibitem{Li} M. Li, {\it Nucl.\ Phys.\ }{\bf B446} (1995) 16.
%
\bibitem{MSS} Y. Matsumura, N. Sakai and T. Sakai, 
{\it Phys.\ Rev.\ }{\bf D52} (1995) 2446.
%
\bibitem{OSS} H. Oda, N. Sakai and T. Sakai, 
preprint  hep-th/9606157 
{\it Phys.\ Rev.\ } in press.
%
%
\bibitem{LeNaTh} F. Lenz, H. W. L. Naus, and M. Thies, 
{\it Ann.\ Phys.\ } (N.Y.) 
{\bf 233} (1994) 317.
%

\end{thebibliography}
\end{document}